\documentclass[twocolumn,aps,prb]{revtex4}

\usepackage{amsmath}
\usepackage{amssymb}
\usepackage{graphicx}
\usepackage{epsfig}
\usepackage{bm}
\usepackage{units}
\usepackage{subfigure}
\usepackage{hyperref}
\usepackage{breakurl}

\newcommand{\be}{\begin{equation}}
\newcommand{\ee}{\end{equation}}
\newcommand{\bea}{\begin{eqnarray}}
\newcommand{\eea}{\end{eqnarray}}
\newcommand{\ben}{\begin{enumerate}}
\newcommand{\een}{\end{enumerate}}
\newcommand{\bit}{\begin{itemize}}
\newcommand{\eit}{\end{itemize}}

\newcommand{\la}[1]{\label{#1}}

\newcommand{\Fig}[1]{Fig.~\ref{#1}}
\newcommand{\Figu}[1]{Figure~\ref{#1}}

\newcommand{\bert}{\raise-0.45mm\hbox{\Large$\Box$}}		

\begin{document}

\title{Isaac Newton's sinister heraldry}
 
\author{Alejandro Jenkins}\email{ajenkins@fisica.ucr.ac.cr}

\affiliation{Escuela de F\'isica, Universidad de Costa Rica, 11501-2060 San Jos\'e, Costa Rica}

\date{Oct.\ 2013, last revised Jul.\ 2014} 

\begin{abstract}

After Isaac Newton was knighted by Queen Anne in 1705 he adopted an unusual coat of arms: a pair of human tibi\ae ~crossed on a black background, like a pirate flag without the skull.  After some general reflections on Newton's monumental scientific achievements and on his enigmatic life, we investigate the story of his coat of arms.  We also discuss how its simple design illustrates the concept of chirality, which would later play an important role in the philosophical arguments about Newton's conception of space, as well as in the development of modern chemistry and particle physics. \\



\end{abstract}

\maketitle


{\footnotesize
\begin{tabbing}
\hspace{0.165 \textwidth}

\= \hskip 1.74 cm The hearts of old gave hands, \\
\> But our new heraldry is hands, not hearts. \\
\` --- {\it Othello}, act 3, scene 4
\end{tabbing}}

\section{Newton, the man}
\la{sec:man}

According to Lagrange, the greatest and most fortunate of mortals had been Isaac Newton (1642--1727), because it is only possible to discover once the system of the world.\cite{Lagrange}  Newtonian mechanics is no longer regarded as the last word on the workings of the Universe, but it seems likely that as long as there continue to be physics students they shall begin by learning Newton's laws of motion.\cite{Weinberg}

Newton's scientific reputation has long been secure, but the details of his personal and intellectual development have undergone much posthumous scrutiny and second-guessing, from which no clear portrait has emerged.  According to theoretical physicist Martin Gutzwiller, ``although many documents concerning Newton's life and work have been discussed at great length [...], he remains a lonely and mysterious figure with achievements to his credit that have no equal in the history of science.''\cite{Gutzwiller}

The iconoclastic mathematician Augustus De Morgan and other 19th-century scientists fought the British scientific establishment's hero-worship of Newton by dwelling upon ---and sometimes exaggerating--- the least attractive aspects of his personality.\cite{Higgitt} Stephen Hawking's vivid diatribe on Newton's ``deviousness and vitriol,'' in his bestselling {\it Brief History of Time},\cite{Hawking} belongs to that tradition.  Even such a great admirer of Newton's creative genius as the Soviet mathematician Vladimir Arnold could not discuss Newton's career without some disparaging commentary on his character and beliefs.\cite{Arnold}

Historian Richard S.~Westfall, who dedicated more than twenty years to his authoritative biography of Newton,\cite{Westfall} later expressed a personal loathing for his subject.\cite{loathing}  To be sure, the picture of Newton that can be gleaned today from the conflicting accounts of those who knew him and from his multifarious personal papers is neither transparently coherent nor conventionally seductive.\cite{Newtoniana}  Computer scientist Ernest Davis has characterized Newton as ``a genius'' and also ``a very strange man, with a very different viewpoint, who lived in a very different world, and who died almost 300 years ago.''\cite{Davis}

Newton was high-strung, wary and secretive in his dealings with others, and intolerant of criticism, but there are few instances in which he can be convicted of bad faith.  Intellectual rivalries, contests, intrigues, and priority disputes were particularly frequent and acrimonious at a time when scientific publication was in its infancy and savants still depended on aristocratic patronage.\cite{Bernoulli}  In his notorious and unedifying disputes with Hooke,\cite{Hooke} Flamsteed,\cite{Flamsteed} and Leibniz,\cite{Leibniz} Newton did have legitimate grievances.  Those controversies generated more heat than others of the period primarily because of the greater significance of the scientific work involved.\cite{Huygens}

In his lifetime, Newton's ``natural philosophy'' withstood many attacks, most of them misguided and some of them malicious.\cite{Feingold}  In the fraught intellectual climate of the period, those controversies threatened to spill into religious and political disputes that could have cost Newton his livelihood and reputation.\cite{antiNewton}  Newton's guardedness in publishing and explaining his discoveries, as well as his prickly defense of his honor when faced with criticism, do not seem quite so peculiar or paranoid in their historical context.  Had he not enjoyed Restoration England's relative freedom of thought and, from an early age, the security of his Cambridge professorship, it is hard to see how his revolutionary work could have been accomplished.

In evaluating Newton's character, one should also bear in mind that during the last thirty years of his life he served as chief officer of the Royal Mint, earning nearly unanimous praise for his diligence and honesty.  According to Mint official and historian Sir John Craig, Newton was not only personally conscientious but also ``a good judge and handler of men,'' endowed with ``magnetism which in many engendered an extraordinary regard and respect.''\cite{Craig}  In the estimation of the eminent economist Lord Keynes, Newton was ``one of the greatest and most efficient of our civil servants.''\cite{Keynes}  Newton's efforts to improve the coinage contributed significantly to the nation's finances.\cite{Belenkiy}  After his appointment to the Mint had made him a wealthy man, he became known to his extended family and to others for his charitable giving.\cite{charity}

\section{Occult studies}
\la{sec:occult}

Keynes had acquired many of Newton's private manuscripts.\cite{papers}  He was therefore aware of the great scientist's interests in both alchemy and theology, leading him to declare in 1946 that Newton had not been ``the first of the age of reason'' but rather ``the last of the magicians.''\cite{Keynes}  Scholars have since debated whether Newton's alchemical and religious pursuits are properly characterized as ``magical'' and how they relate to his work in physics and mathematics.

The mature Newton showed no interest in astrology,\cite{astrology} communication with spirits, ritual magic, or other of the common occult pursuits of the day.\cite{magic}  The editor of Newton's mathematical papers, D.~T.~Whiteside, could find ``no hint of any belief by him in number-mysticism,''\cite{Whiteside} while recent scholarship has documented Newton's hostility to neo-Platonism, Gnosticism, Kabbalism and other esoteric traditions influential in the Renaissance.\cite{Goldish}

Newton's private writings do evince a keen involvement with alchemy and its arcane imagery, but at the time there was no clear boundary between alchemy as mysticism and alchemy as early modern chemistry.  Moreover, the otherworldliness of alchemy as practiced in Newton's day may have been exaggerated by 19th-century occultists and those influenced by their interpretations.\cite{Principe}  Newton's alchemical interests, which he shared with eminent contemporaries such as Robert Boyle and John Locke, undoubtedly influenced his atomism.\cite{Newman}  On the other hand, the notion that alchemy contributed to his conception of gravity as an action at a distance may be, according to the most recent scholarship, ``something of a canard.''\cite{canard}

Leibniz, who rejected the Newtonian theory of gravity, originated the claim that Newton conceived of gravity as an ``occult quality,'' an aspersion that Newton strenuously denied.\cite{Janiak}  Explaining that he did not ``feign hypotheses'' about the mechanism by which gravity acted across empty space, Newton presented his theory as a mathematical description of gravity's observable effects, an attitude wholly compatible with modern scientific principles.\cite{modern}  According to a tradition collected by Voltaire, when the elderly Newton was asked how he had discovered his celebrated law of universal gravitation, he replied: ``by thinking on it continually.''\cite{thought}

\section{Religion}
\la{sec:religion}

\begin{figure} [t]
\begin{center}
	\includegraphics[width=0.15 \textwidth]{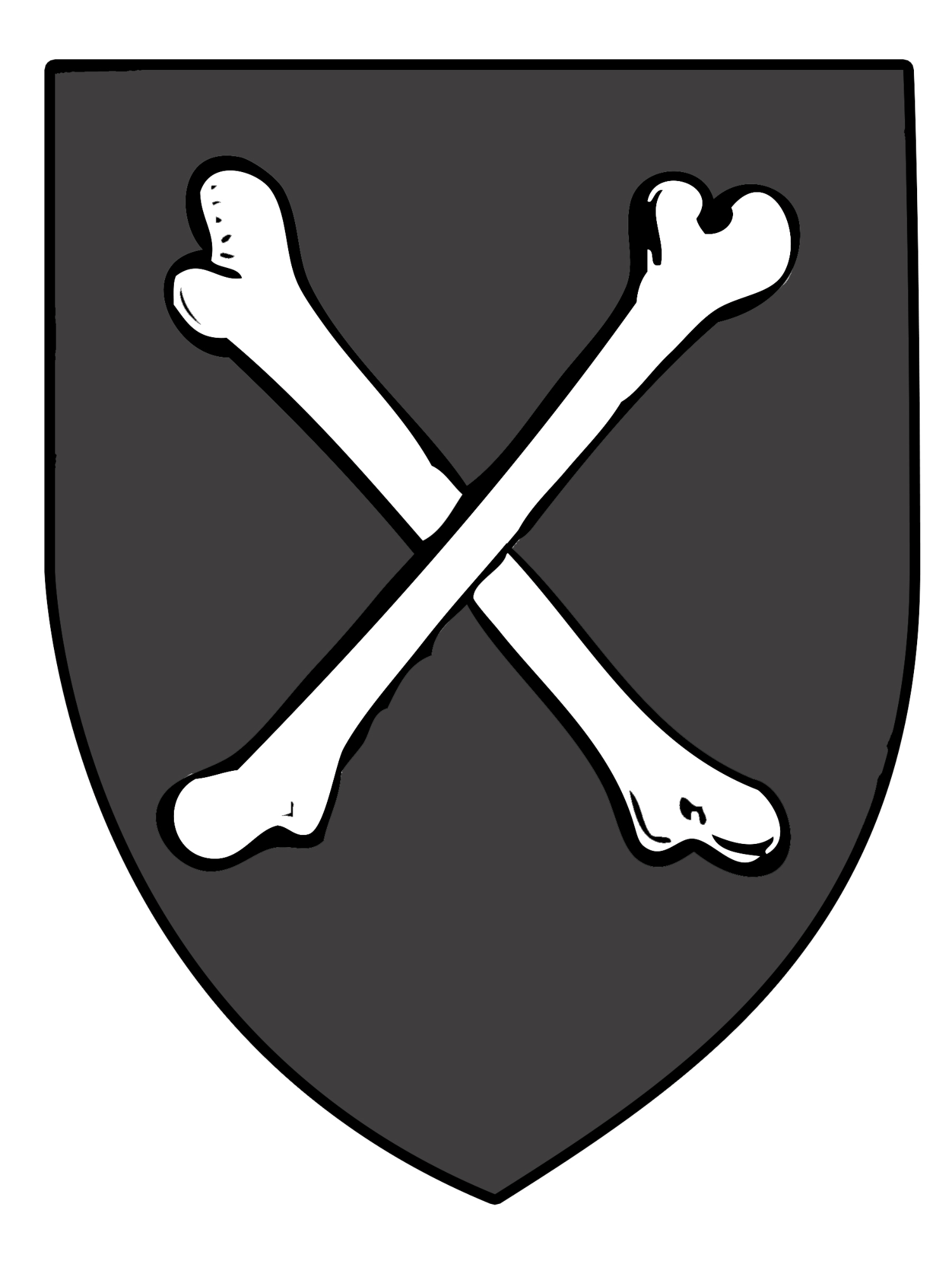}
\end{center}
\caption{\small  Coat of arms of Sir Isaac Newton (1642--1727)\la{fig:Cambridge}}
\end{figure}

Newton was a devout but unorthodox Christian who privately rejected the doctrines of the Holy Trinity\cite{Wiles} and the immortality of the soul as unscriptural and idolatrous.\cite{Nicodemite}  He dedicated a considerable effort to establishing a chronology of human civilization, combining his belief in the wisdom of the ancients and in the historicity of the Bible with innovative uses of astronomy and other quantitative techniques.\cite{chronology}  A failure from the standpoint of our current understanding of history, that work does shed light on Newton's attitudes towards the various branches of learning.

Philosopher Richard Popkin tried to make theology the key to Newton's intellectual development, provocatively asking why ``one of the greatest anti-Trinitarian theologians of the 17th century'' took ``time off to write works on natural science, like the {\it Principia Mathematica}?''\cite{Popkin}  But the fact that, in due course, Newton's mathematics and physics were decisively embraced by an international community of scholars, the vast majority of whom were either ignorant of his religious convictions or actively hostile to them, suggests that Newton's theology can contribute little to elucidating the {\it content} of his science.\cite{Euler}  Even Newton's belief that divine intervention is needed to keep the planets on their regular orbits ---as theologically charged a claim about nature as he ever published\cite{Opticks}--- was based on the mathematical laws of gravity that he had abstracted from astronomical observations, and the problem that he thereby raised, the dynamical stability of the solar system, has continued to occupy physicists and mathematicians to this day.\cite{Tremaine}

In my view, if there is something to learn today from Newton's religion, and more broadly from his life beyond his immortal work in the exact sciences, it is a lesson close to Max Weber's ``elective affinities'' between the worldview associated with certain historical strands of Protestant Christianity, and the scientific and industrial revolutions.\cite{Merton}  According to historian Stephen Snobelen, ``in his biblicism, piety and morality, Newton was a puritan through and through.''\cite{Nicodemite}  The central paradox of Newton's life is close to what Weber underlined in his analysis of the growth of capitalism: that the Puritans, whose convictions seem so remote from our perspective, contributed decisively to creating the modern world.\cite{Weber}  In the context of the radical English Puritanism of the 17th century ---of the religion of Oliver Cromwell, John Milton, and the Pilgrim Fathers--- the man Isaac Newton gains some intelligibility.\cite{Unitarianism}

Puritanism and its legacy have long been contentious in English-language historiography, as reflected, for instance, in the widely diverging evaluations of Oliver Cromwell's rule.\cite{Cromwell}  Newton's own (doctrinally heterodox) Puritanism may account, in part, for the antipathy of some of his biographers.  Historian Frank Manuel saw the prosecution of counterfeiters during Newton's tenure at the Mint as an opportunity for the great man to ``hurt and kill without doing violence to his scrupulous puritan conscience,'' adding that ``the blood of the coiners and clippers nourished him.''\cite{Manuel}  Rupert Hall dismissed this ``sadistic vampire'' portrait as ``blood-tub Victorian melodrama rather than biography.''\cite{vampire}  

\section{Knighthood and genealogy}
\la{sec:knight}

\begin{figure} [t]
\begin{center}
	\includegraphics[width=0.25 \textwidth]{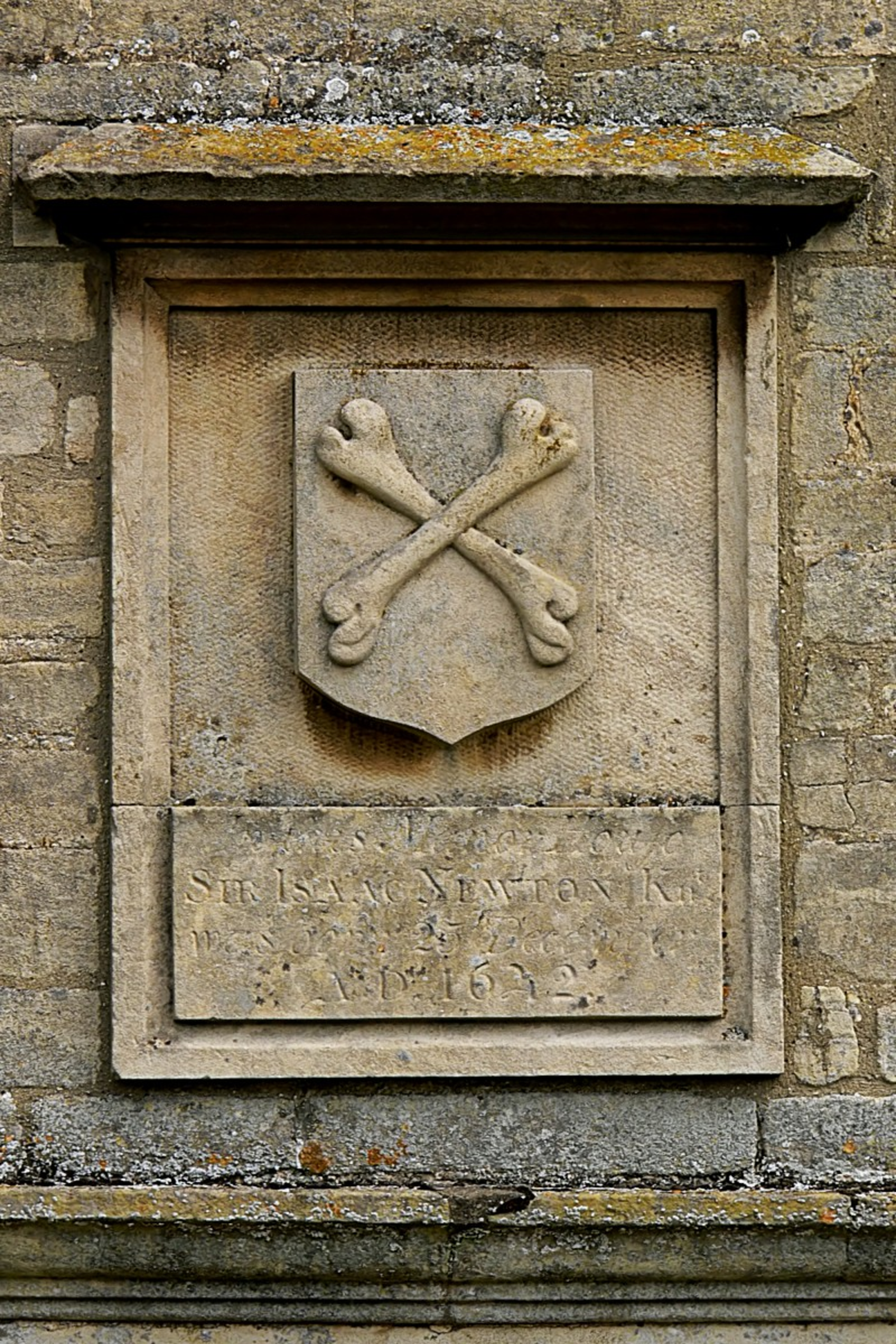}
\end{center}
\caption{\small Stone tablet above the front door of Woolsthorpe Manor, near Colsterworth, in Lincolnshire.  The inscription reads ``In this Manor House Sir Isaac Newton, Knt.\ was born, 25th December, A.D. 1642.''  The picture is by Flickr user Walwyn,\cite{Walwyn} and is used here under the terms of the Creative Commons 2.0 license.\la{fig:Woolsthorpe}}
\end{figure}

On 16 April 1705, in a ceremony conducted at Trinity College, Cambridge, Queen Anne elevated Isaac Newton to the rank of knight bachelor.\cite{Bacon}  At the time, that honor was usually granted to military officers and senior figures in the national and local governments, as well as to rich merchants and others with political connections.\cite{LeNeve}  Newton's knighthood was probably a royal favor to his political patron, Lord Halifax.  Halifax, a prominent figure in the Whig political party, hoped thereby to promote Newton's candidacy in the upcoming parliamentary election, but Newton came last at the polls and never again sought elective office.\cite{knight}

The lawyer James Montagu (Lord Halifax's brother and a future Attorney General) and the academic John Ellis (master of Gonville and Caius College and vice-chancellor of the University of Cambridge) were knighted along with Newton.  More than a few ``new men'' of obscure ancestry, like Ellis, figure among the knights bachelor created by Queen Anne.\cite{LeNeve}  For his part, Sir Isaac soon sought to establish his status as a gentleman.\cite{Heralds}  In November he submitted to the College of Heralds an affidavit detailing his ancestry and claiming Sir John Newton, Bart.,\cite{Bart} a rich landowner from Culverthorpe, in Lincolnshire, as a third cousin.  This was supported by a declaration signed by Sir John.\cite{Foster}

That genealogy might seem suspicious, since Isaac Newton's father had been a relatively prosperous but illiterate yeoman farmer.\cite{yeoman}  It was not uncommon, in that day and age, for a wealthy man to pay an accommodating officer for the right to bear a coat of arms, or to some other hereditary privilege, based on a fictional pedigree.  In fact, the second Sir John Newton, Bart.\ (father of the cousin who supported Isaac Newton's affidavit) had inherited his baronetcy from a Sir John Newton to whom he was unrelated.  The first Sir John, a native of Gloucestershire, was childless.  He agreed to pass on his baronetcy to his rich namesake and putative cousin, resident in Lincolnshire, in exchange for help settling his financial obligations.  But archival research has established that the genealogy in Isaac Newton's 1705 affidavit is genuine and that he prepared it from the available records with his habitual scrupulousness.\cite{Foster,Baird}

According to accounts collected by Sir David Brewster, later in his life Newton claimed in private conversation that his male-line ancestry might have been Scottish.\cite{Scottish}  Yet it seems quite unlikely that Newton could have seriously entertained that idea, which contradicts the well-documented genealogy that he had submitted to the College of Heralds.  Moreover, it is clear from their mutual correspondence that Isaac Newton consistently regarded Sir John Newton as head of their common clan.  Sir John's heir, the rich and politically connected Sir Michael Newton, would serve as chief mourner at Isaac Newton's funeral in 1727.\cite{clan}

\section{Blazonry}
\la{sec:blazon}

\begin{figure} [t]
\begin{center}
	\includegraphics[width=0.15 \textwidth]{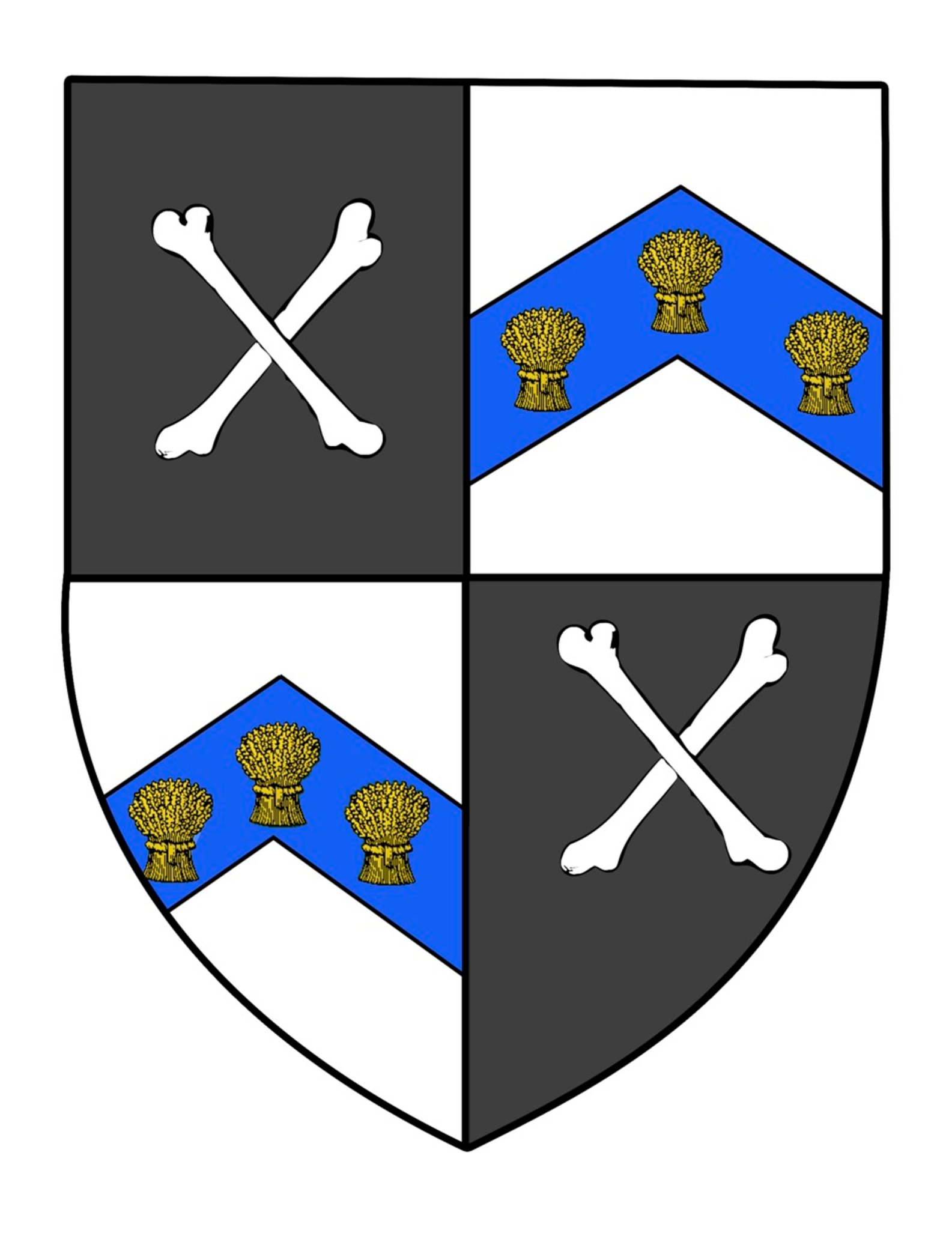}
\end{center}
\caption{\small Escutcheon of Sir Michael Newton ({\it ca}. 1695--1743), 4th baronet of Barrs Court in the County of Gloucester (actually resident in Culverthorpe, Lincolnshire), Knight of the Bath, Member of Parliament, as well as Sir Isaac Newton's kinsman and chief mourner.  The crossbones are quartered with another design, blazoned as ``argent, on a chevron azure three garbs or,'' taken from the Gloucestershire family (originally surnamed Cradock) of Sir John Newton, 1st baronet.\cite{Wotton}\la{fig:Barrs}}
\end{figure}

Isaac Newton's personal coat of arms is shown in \Fig{fig:Cambridge}.  This escutcheon is carved on a stone tablet above the front door of Woolsthorpe Manor, the home in Woolsthorpe-by-Colsterworth, near Grantham, Lincolnshire, where Newton was born and where he made some of his greatest intellectual breakthroughs after returning there when the University of Cambridge was closed due to the Great Plague of 1665--66.  That tablet, shown in \Fig{fig:Woolsthorpe}, was installed in 1798 by Edmund Turnor, whose father had purchased Woolsthorpe Manor in 1733.\cite{Manor}

In the language of heraldry, \Fig{fig:Cambridge} is described (``blazoned'') as ``sable, two shinbones in saltire, the dexter surmounted of the sinister argent.''  This requires some explanation for the uninitiated:  ``Sable'' means black and refers to the background color.  A ``saltire'' is an X-shaped cross.  ``Argent'' means white and identifies to the color of the bones.  The ``sinister'' bone (which runs from the upper right to the lower left) surmounts the ``dexter'' (running from the upper left to the lower right).

The last detail is interesting.  The reader may care to verify that the crossbones of \Fig{fig:Cambridge} are arranged like the blades of a pair of left-handed scissors.  Even with symmetrical handles, ordinary (i.e., right-handed) scissors cannot be turned so that they will be congruent with \Fig{fig:Cambridge}: they will always look like the crossbones in \Fig{fig:Barrs}.  The design of \Fig{fig:Cambridge} is therefore unequivocally left-handed, or, to use the Latin term, {\it sinister}.

The blazon to which Isaac Newton was entitled by virtue of the genealogy in the 1705 affidavit did not specify which bone should lie on top.\cite{Foster}  The arms of the Newton baronets show the dexter bone surmounting the sinister (see \Fig{fig:Barrs}).\cite{Wotton}  On the other hand, Samuel Newton (1628--1718), mayor of Cambridge, who was unrelated to Sir Isaac but with whom Sir Isaac must have become acquainted during his years at the University, displayed his own coat of arms with the sinister bone surmounting the dexter.\cite{Cambridge}

The coat of arms of Isaac Barrow, Newton's mentor and predecessor as Lucasian Professor of Mathematics, also features a sinister saltire, in his case formed by two crossed swords.  An ornate carving of it, by the eminent artist Grinling Gibbons, graces the end of a bookshelf in Trinity College's Wren Library, built while Barrow was master and Newton a fellow of that college.\cite{Barrow}

\section{Symbolism}
\la{sec:symbolism}

Like other aspects of his life, Newton's heraldry resists easy interpretation.  Newton did not create the crossbone device, which is ancestral.  On the other hand, one might attribute some meaning to his adoption of such a stark design (which only a couple of decades later would become indelibly associated with the Jolly Roger by the pseudonymous Captain Charles Johnson's {\it General History of Pyrates}\cite{Pyrates}), rather than seeking a new grant of arms from the College of Heralds (which he might well have obtained, given his eminence).

Newton's efforts to uncover factual accounts of history and natural science concealed in ancient sources, such as the Bible, Greek myths, or the {\it Corpus Hermeticum}, were surely related to the ``emblematic worldview'' of other Renaissance intellectuals who sought wisdom by deciphering and correlating the symbolisms that they believed lay hidden in all things.  But, as Matt Goldish has underlined, ``Newton's relationship to symbols was not mystical, like that of the Rosicrucians or other Hermetics, but rather mathematical: he regarded them as a puzzle for which available clues provide a logical solution.''\cite{Goldish-emblems}

In his study of ancient sources, Newton always sought precise, rational meanings that he thought had been obfuscated after the corruption of mankind's first religion.\cite{Hall-Hermetic}  The alchemical literature of his day, with its deliberate encoding of chemical recipes in riddling allegories,\cite{Starkey} must have encouraged that approach.  Scholars who have taken seriously the characterization of Newton as ``last of the magicians'' have tended to downplay this {\it disenchanting} tendency, as well as his consistent rejection of metaphysical speculation divorced from rigorous experimental knowledge.\cite{Levitin}

With this in mind, three non-exclusive possibilities for Newton's interpretation of his own escutcheon appear plausible: as a conventional mark of his social rank and family connections, as a stark {\it memento mori} like the skulls and hourglasses of Puritan gravestones,\cite{Southey} or as a mathematical symbol connected to what would much later be called {\it chirality}.  Though wholly conjectural, the last possibility is intriguing in light of other intellectual developments.

\section{Chirality}
\la{sec:chirality}

An object is said to be chiral if it cannot be superimposed on its mirror image.  The term, coined by Lord Kelvin in 1893,\cite{etymology} is taken from the Greek word for hand ($\chi \varepsilon \acute{\iota} \rho$), the human hands being familiar instances of chiral objects: the left and right hands are mirror images of each other, and a left-hand glove will not fit over a right hand, or vice-versa.  Some 19th-century mathematical scientists, including Clerk Maxwell and Willard Gibbs, referred to the geometrical transformation that replaces a figure with its mirror image as a ``perversion,''\cite{Gibbs} a usage that is now largely obsolete but persists in some scientific contexts.\cite{perversion}

Awareness of the incongruence of mirror images is much older.  Modern chemists often credit Ren\'e Descartes (1596--1650) with this insight, based on the remark attributed to him that ``any man who, upon looking down at his bare feet, does not laugh, has either no sense of symmetry or no sense of humor.''\cite{feet}  I have been unable to find this quote in Descartes's published works or correspondence, but Descartes did have a sophisticated understanding of what would later be termed chirality, for he famously explained magnetism as a flow of tiny particles shaped as left-handed or right-handed corkscrews, whose mutual action could therefore produce either attraction of repulsion.\cite{screws}  Newton rejected such {\it ad hoc}, unverifiable hypotheses as ``fictions.''\cite{DeGravitatione}  But when it came to the double refraction (now called birefringence) exhibited by crystals of Iceland spar, Newton proposed an interpretation in terms of the ``sides'' of light.\cite{birefringence,matter} (These ``sides,'' akin to the linear polarizations of modern physical optics, are not chiral.)

Did Descartes's interest in the incongruence of mirror images influence Newton?  Though I have found no direct evidence of this, the possibility is not far-fetched.  Newton was both Descartes's greatest disciple and his greatest critic.  In the words of mathematician and science historian Clifford Truesdell, ``if the steely-splendid preface to the {\it Principia} is not just a bolt from the blue (and where in science as such bolts to be found?), who else than Descartes can be its grandsire?''\cite{steely}

\subsection{Rightful delineation}
\la{sec:rightful}

The young Newton first came to the attention of the ``Republic of Letters'' as creator of the first practical reflecting telescope.\cite{telescope}  An issue that might have caught Newton's attention is how a telescope can alter an image's orientation.   Galileo's old refracting telescope leaves orientation unchanged, but Kepler's improved refractor rotates it by 180$^\circ$.  The image can be made upright by reflection (a desirable feature when observing objects on land), at the cost of reversing left and right.  Newton's telescope used a pair of mirrors, one curved, the other flat.  Each mirror reverses chirality, but the net effect is that the image is rotated while chirality is preserved.

In his later years, Newton had further occasion to consider the incongruence of mirror images when he sought to estimate the date of the expedition of the Argonauts from what he took to be a contemporaneous description of the positions of the equinoxes and solstices with respect to the constellations.  In the course of that work, Newton had to correct for inconsistencies in the chiralities of constellations as drawn in stellar atlases.  He referred to the corrected stellar maps as being ``rightly delineated,'' confusing some of his critics.\cite{constellations}  The reason for those inconsistencies is that constellations may be depicted as they are actually seen from the Earth or as they would appear on the surface of a celestial globe (i.e., from an imaginary vantage point outside the celestial sphere).\cite{GrandCentral}

\subsection{Philosophy}
\la{sec:philosophy}

\begin{figure} [t]
\begin{center}
	\includegraphics[width=0.2 \textwidth]{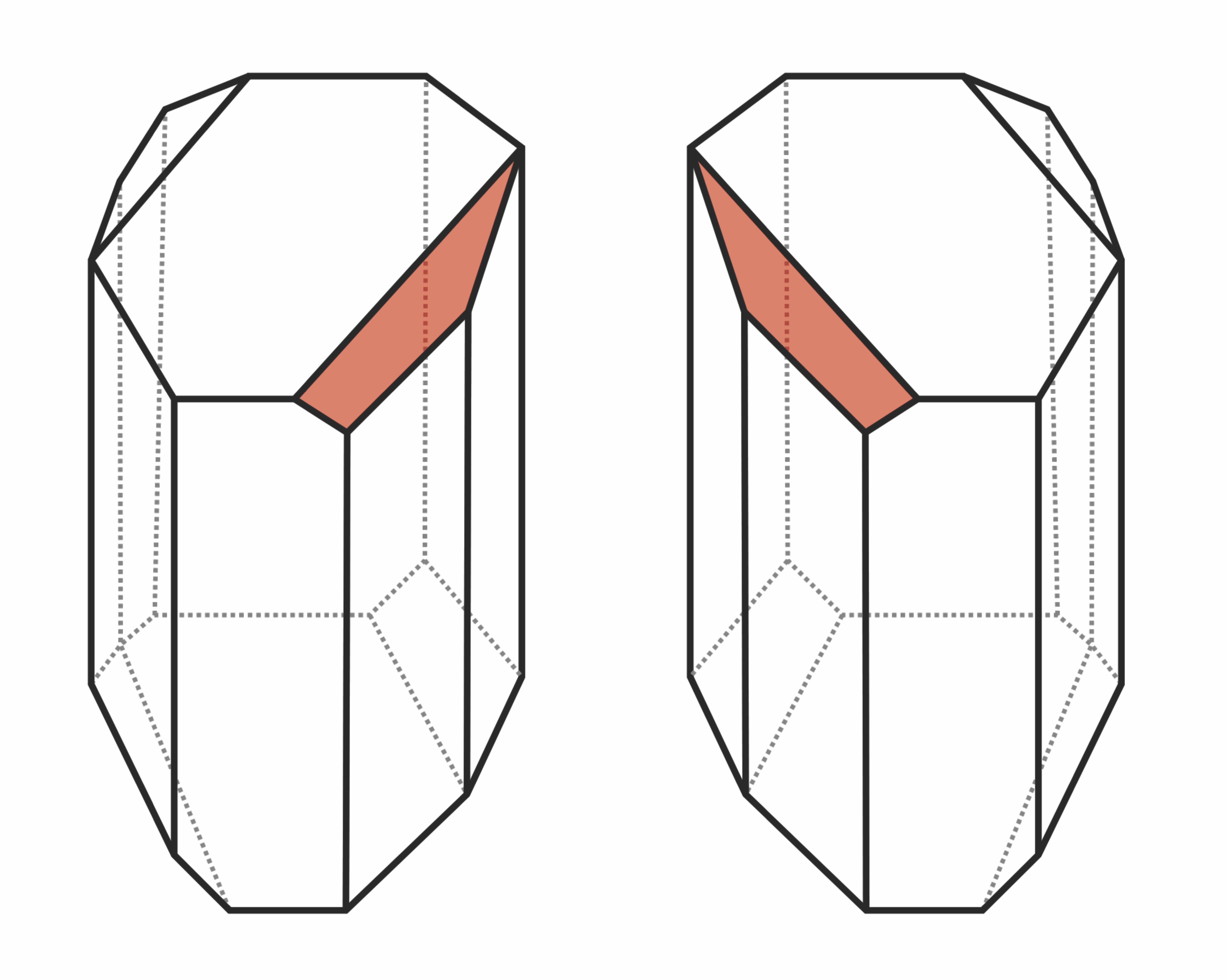}
\end{center}
\caption{\small Idealized shapes of the crystals of the two chiral forms of sodium ammonium tartrate, which Pasteur observed in 1848.  One of the minor facets has been colored to make the chirality more obvious.  Image from Wikimedia Commons, available at \url{http://commons.wikimedia.org/wiki/File:Pcrystals.svg}\la{fig:tartrate}}
\end{figure}

In 1768, Immanuel Kant considered chirality (which he called the problem of ``incongruent counterparts'') in the context of the philosophical debate on whether the concept of space is reducible to the relations between concrete objects.\cite{Kant1}  This was a question that dated back to the disputes between Newton and Leibniz.\cite{Alexander}  Kant argued that the spatial relations between the fingers of a given hand are the same whether it be a left or a right hand and that the fact that the two hands are distinguishable must therefore be explained by appeal to an external geometrical space in which the hands exist.  According to Kant, this supported Newton's belief in the existence of an absolute space, which Leibniz had denied.\cite{Kant2}

Chirality figures also in Ludwig Wittgenstein's {\it Tractatus Logico-Philosophicus} of 1921.  Wittgenstein pointed out that a right-hand glove could be worn over a left hand if the three-dimensional glove could be turned around in a fourth spatial dimension.\cite{Wittgenstein}  (An equivalent observation had been made by mathematician August M\"obius in 1827.\cite{Mobius})  Wittgenstein's comment is characteristically gnomic, but he seems to have concluded from this that chirality is tied to the space in which the object is embedded (and therefore to the ways in which it is allowed to move) rather than being a property of the object itself.\cite{vanCleve}

\subsection{Chemistry}
\la{sec:chemistry}

\begin{figure*} [t]
\begin{center}
	\includegraphics[width=0.6 \textwidth]{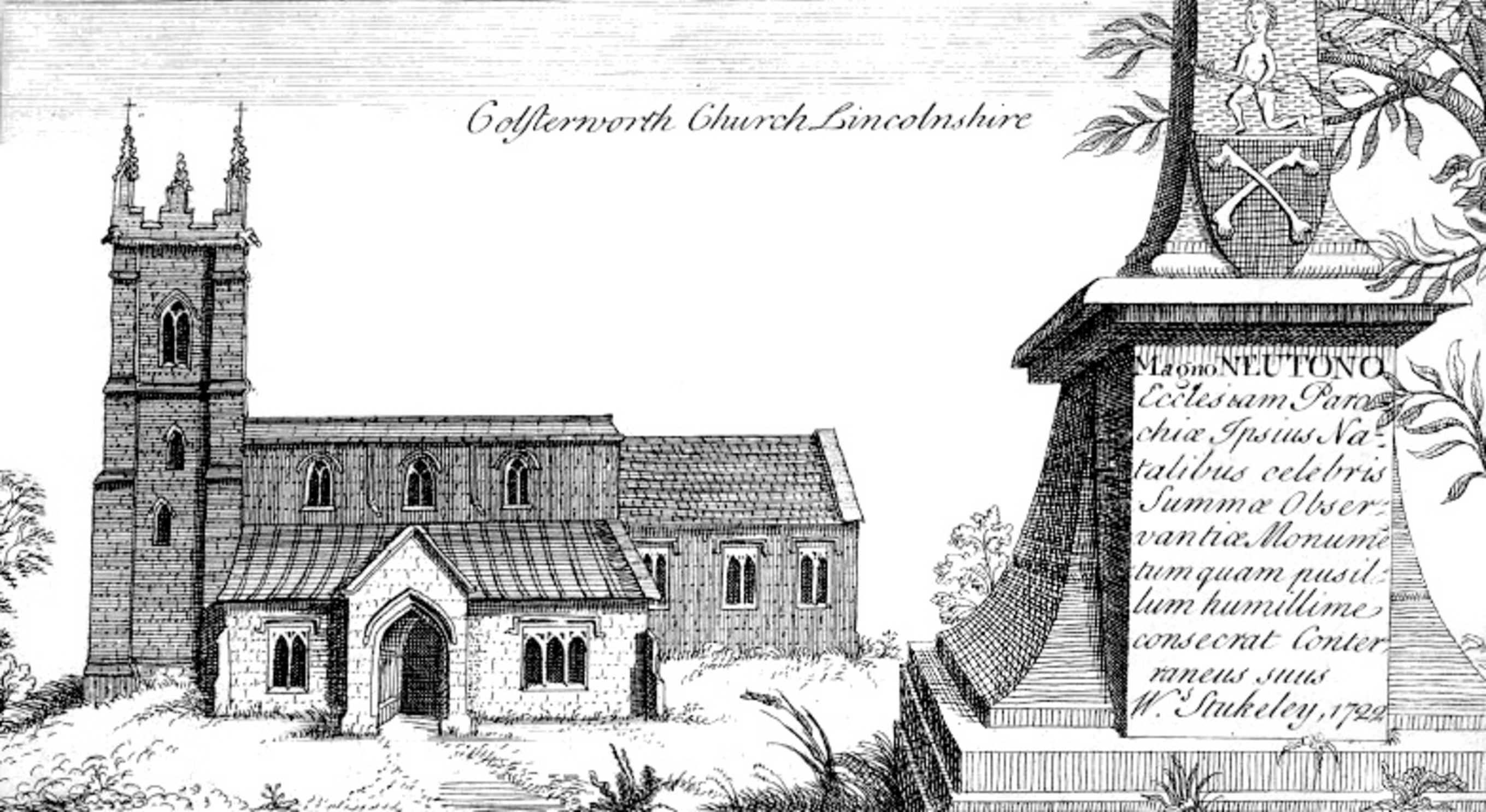}
\end{center}
\caption{\small Engraving by William Stukeley, dated 1722, of the Church of St.\ John the Baptist, Colsterworth, where Isaac Newton was baptized.  The monument on the right is an imaginary decoration.  Image from Wikimedia Commons, available at \url{http://commons.wikimedia.org/wiki/File:Colsterworth_Church,_Lincolnshire_by_W._Stukeley_(1722).png}\la{fig:Stukeley}}
\end{figure*}

The young Louis Pasteur, who had just completed his doctorate in chemistry and physics, was faced with a puzzle.  He knew that $(+)$-tartaric acid, extracted from fermented grape juice, is optically active (meaning that it rotates the plane of the polarization of light passing through it).  Factory-made ``paratartaric acid'' (now called racemic tartaric acid) has the same elemental composition as $(+)$-tartaric acid, yet crystallizes differently and is optically inactive.  In 1848, Pasteur discovered that the artificially synthesized acid could make two distinct crystals, shown in \Fig{fig:tartrate}, which are mirror images of each other.  By carefully picking out crystals of one form and dissolving them, he produced a substance identical to the natural $(+)$-tartaric acid.\cite{Pasteur}

Pasteur's work established that racemic tartaric acid is a mixture of two substances whose molecules are mirror images of each other (``enantiomers''), and that only one of the two forms occurs in nature.  A decade afterwards, his experiments on tartrate fermentation led Pasteur to conclude that chirality is an essential factor in molecular recognition within biological systems.\cite{Gal} Pasteur's brilliant elucidation of molecular chirality would play a key role in the growth of biochemistry. Writing a century later, the great crystallographer and molecular biologist J.~D.~Bernal described it as Pasteur's ``first and in some ways his greatest scientific discovery.''\cite{Bernal}

Pasteur saw deep significance in the homochirality of biomolecules, writing that ``life as manifested to us is a consequence of the dissymmetry of the Universe.''\cite{Pasteur-life}  Why life on Earth evolved to use only one enantiomer of such organic substances as amino acids and sugars remains mysterious.\cite{homochirality}

\subsection{Particle physics}
\la{sec:particle}

In 1956, theoretical physicists T.~D.~Lee and C.~N.~Yang proposed that the weak nuclear interaction, unlike the other forces of nature, could distinguish between mirror images.  Soon afterwards, C.~S.~Wu confirmed this experimentally when she found that in radioactive decays of cobalt-60 nuclei the electron produced is more likely to be emitted in a direction opposite to the nucleus's intrinsic angular momentum (``spin''), making the decay process left-handed.\cite{parity}

A measurement of the spin of a massless particle along the direction of its linear momentum has only two possible outcomes, which are mirror images of each other and are therefore called chiralities.  For the photon (the massless particle of light), these correspond to its two circular polarizations.\cite{polarization}  Elementary particles that do have mass (such as the electron) are now regarded as mixtures of the two chiralities.\cite{helicity}

The weak interaction acts exclusively on the left-handed component of particles and on the right-handed component of antiparticles.  The Standard Model (SM) of high energy physics is therefore formulated in terms of chiral quantum fields.\cite{SM}  In 1964, Cronin and Fitch reported experimental evidence that there is also a slight asymmetry between the interactions of left-handed particles and right-handed antiparticles.  In 1973, Kobayashi and Maskawa showed how this effect, known as ``charge-parity'' (CP) violation, was possible within the theoretical context of the SM.  The amount of CP violation in the SM is, however, far too small to explain why the Universe contains as much matter as it does, but no antimatter.  How CP violation in the early Universe can have been large enough to account for the matter content that we see today is one of the great unsolved problems of both cosmology and particle physics.\cite{CP}

\section{The armigerous Newton}
\la{sec:armiger}

\begin{figure*} [t]
\begin{center}
	\includegraphics[width=0.55 \textwidth]{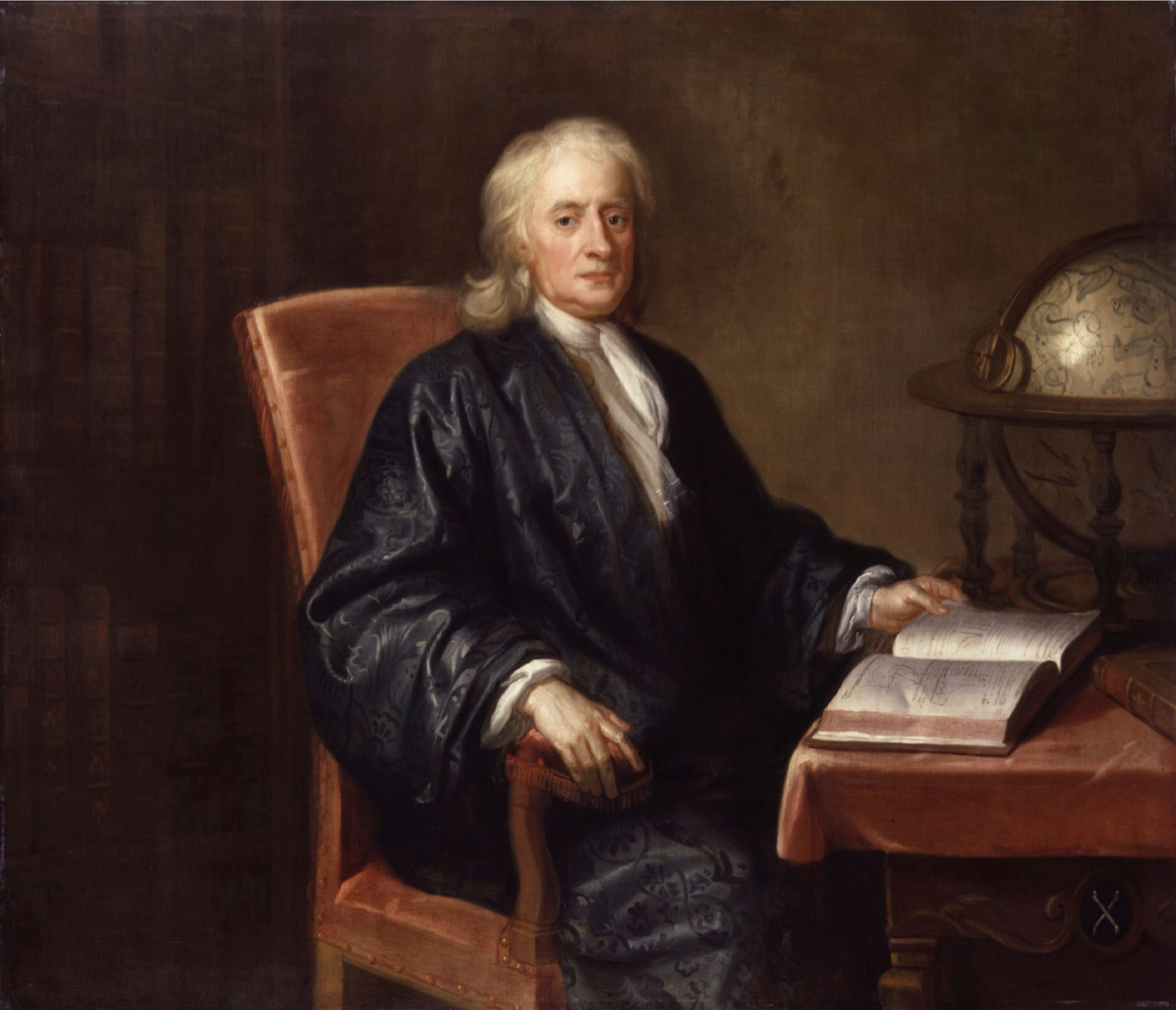}
\end{center}
\caption{\small Portrait of Sir Isaac Newton by the studio of Enoch Seeman, painted {\it circa} 1726.  Note the small heraldic decoration on the side of the table.  Copyright of the National Portrait Gallery, London, used here with permission.\la{fig:Seeman}}
\end{figure*}

Like other wealthy armigers of the period, Sir Isaac had his coat of arms painted on the doors of his carriage.  A French commentator mistook it for a ``death's head'' and interpreted it as evidence that Newton ---whose Christian convictions were problematic for the free-thinking {\it philosophes} who embraced his science--- had ``entirely taken to religion'' in his old age.\cite{carriage}  (Laplace and others would later pursue the idea that Newton's religiosity was a symptom of senility or derangement, perhaps subsequent to his nervous breakdown of 1693.\cite{Laplace-religion})

Members of other English families surnamed Newton also bore crossbones on their coats of arms.\cite{Burke}  That heraldic device appears to be medieval and might have been intended originally as a symbol of warlike prowess.  Several armigerous Newtons (including the Newton baronets) displayed, as a crest above the shield with the crossbones, the image of an ``eastern prince'' (or in some cases a ``naked man'') kneeling on his left knee and surrendering his sword.\cite{Burke}  According to a family tradition, Sir Ancel Gorney, an ancestor of the Newtons, fought along King Richard I in the Third Crusade and captured a Muslim prince at Ascalon.\cite{Wotton}

In 1722, William Stukeley, a personal friend of Sir Isaac and a noted antiquarian, engraved a drawing of the church where Newton was baptized (see \Fig{fig:Stukeley}), and later included it in his manuscript biography of Newton.\cite{Stukeley}  That illustration incorporates, as a sort of caption, an imaginary monument with a heraldic decoration and a Latin inscription.  The kneeling figure in the crest suggests that Stukeley based his decoration on the heraldry of the Newton baronets.\cite{Colsterworth}

The coat of arms appears as well on the portrait shown in \Fig{fig:Seeman}, made in the final years of Newton's life.  The use of the heraldic device suggests that this particular painting ---closely patterned after an earlier one by Enoch Seeman--- was made for the Newton family.\cite{icon}  Note that it depicts the dexter chirality for the crossbones, contrary to what is shown in Figs.~\ref{fig:Cambridge} and \ref{fig:Woolsthorpe}.  The dexter chirality is also used in the arms carved by John Woodward in 1755--56 above the stalls in the chapel of Trinity College.\cite{Trinity}

\section{Bend sinister}
\la{sec:sinister}

The fictional protagonist of Neal Stephenson's 2003 novel {\it Quicksilver} visits Newton at Woolsthorpe Manor during the Great Plague and anachronistically notes the crossbones of \Fig{fig:Woolsthorpe}, a device whose ``awfulness'' embarrasses him as an Englishman.\cite{Quicksilver}  In fact, despite living in an age when heraldry was highly prized as a mark of identity and status, Newton left a tenuous heraldic trail.  I have been unable to confirm that he personally chose, or even used, the sinister chirality illustrated in \Fig{fig:Cambridge}.\cite{Noel}  As detailed above, the relevant documents from the College of Heralds specify no chirality, while surviving instances in which a coat of arms is associated with Isaac Newton generally show the dexter chirality of the Newton baronets, with the significant exception of the tablet pictured in \Fig{fig:Woolsthorpe}.  The sinister chirality was associated with the Newtons of Cambridge, whom Isaac Newton must have known but to whom he was unrelated.

The genteel Turnor family of Stoke Rochford assiduously cultivated the memory of their illustrious countryman.  \Figu{fig:Roubillac} shows their copy of the bust of Newton by sculptor Louis Fran\c{c}ois Roubillac, whose original resides in the Trinity College library.  Antiquarian Edmund Turnor (1754--1829),\cite{Turnor-DNB} grandson of the man of the same name who purchased Woolsthorpe Manor in 1733, recorded only the dexter chirality as connected with Sir Isaac's family.\cite{Turnor}  But his father (another Edmund) may have had reason to install such a permanent and conspicuous depiction of the sinister chirality at the most significant of all locales associated with Newton's life.\cite{notablet}

In heraldry, a diagonal band running from the upper left-hand to the lower right-hand corner is called a ``bend.''  The mirror image, with the band running from the upper right-hand to the lower left-hand corner, is identified as a ``bend sinister.''  The reason for this terminology is that if the coat of arms were drawn on the front of a battle shield, the top of the bend sinister would be near the knight's left shoulder.  Was Newton aware that in the English heraldry of his day a bend sinister often marked the arms of a bastard child?\cite{bastard}

\section{Last wonderchild}
\la{sec:wonderchild}

\begin{figure} [t]
\begin{center}
	\includegraphics[width=0.2 \textwidth]{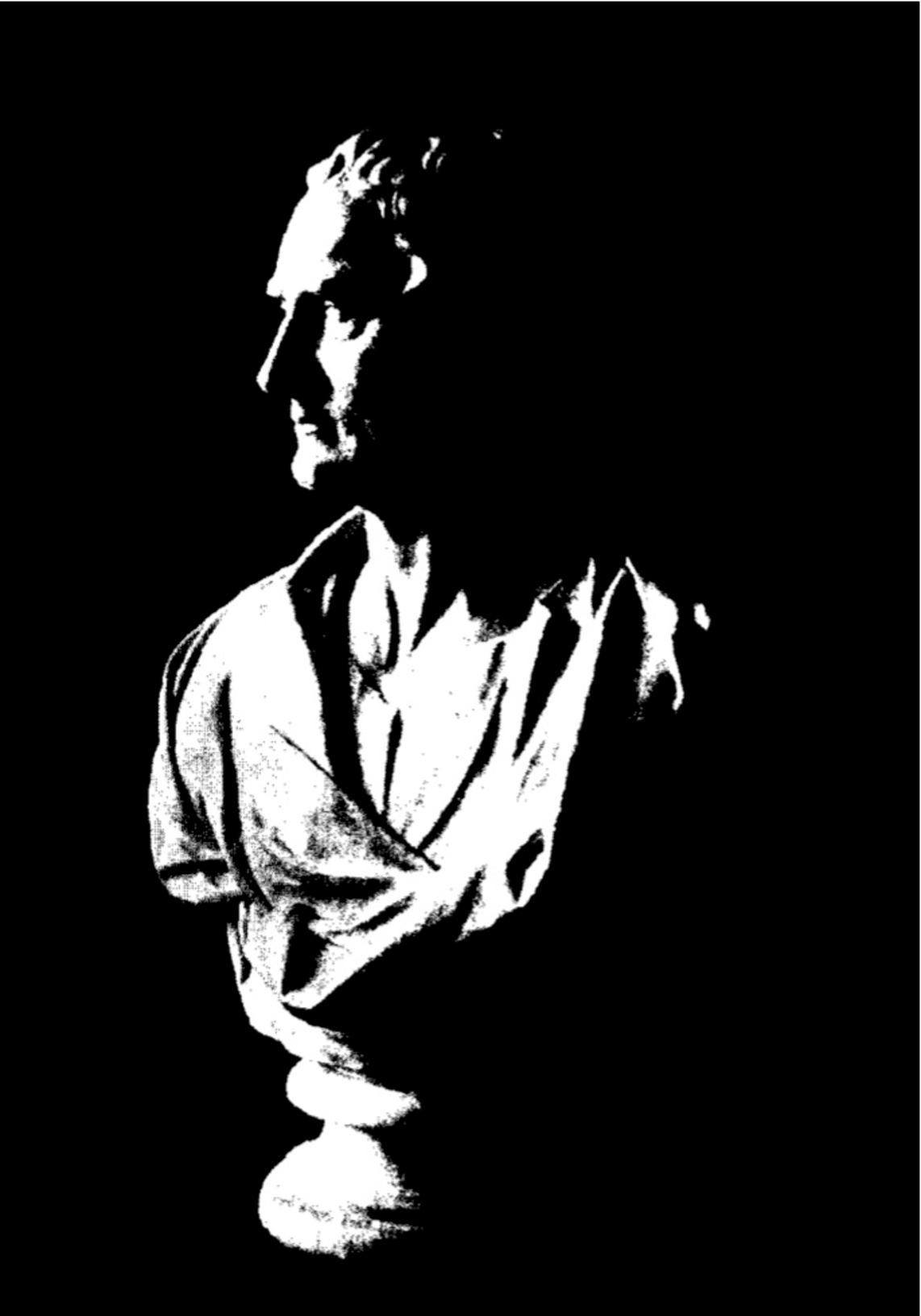}
\end{center}
\caption{\small Isaac Newton, from a copy of a bust by Roubillac owned by the Turnor family.  The picture is taken from Ref.~\onlinecite{Foster} and is used here with permission of the Society for Lincolnshire History and Archaeology.\la{fig:Roubillac}}
\end{figure}

Newton's father had died three months before his birth.  When his mother remarried, the three-year-old Newton was left in the care of his maternal grandparents.  Treated almost as an orphan, Newton had a lonely and unhappy childhood, in which some biographers have seen a source of his adult neuroses.  In Keynes's view, ``Isaac Newton, a posthumous [son] born with no father on Christmas Day,\cite{Gregorian} 1642, was the last wonderchild to whom the Magi could do sincere and appropriate homage.''\cite{Keynes}

It is easy to miss Newton's personal dimensions or to mistake them altogether.  Like everyone else, he had a genealogy, both familial and intellectual.  He was a creature of his own time and place, his work a part of what Einstein called ``the doubtful striving and suffering of his generation.''\cite{Einstein}  Yet he was also a prime mover in what may be the grandest cultural shift in history.  He left no biological descendants, but he is an ancestor to modern humankind.  Neither the standards of his day nor those of ours are quite apt for giving his full measure.  And neither the hero-worship of his first biographers nor the various revisionisms of subsequent scholars have produced a compelling picture of how such a man was possible.\cite{Glashow}

On his deathbed, Newton refused the last rites of the Church of England.  He died on the morning of 20 March 1727.  A week later his body lay in state in the Jerusalem Chamber at London's Westminster Abbey.  On 4 April the Lord Chancellor and five other noblemen carried the remains to their resting place beneath the Abbey's nave, under a gravestone marked {\it Hic depositum est quod mortale fuit Isaaci Newtoni} (``here lies what was mortal of Isaac Newton'').

\begin{acknowledgements}

I thank Neer Asherie and Joseph Gal for correspondence on Pasteur's work, for bringing Ref.~\onlinecite{Gal} to my attention, and for suggestions on improving this manuscript.  Jed Buchwald and Mordechai Feingold pointed out Refs.~\onlinecite{carriage} and \onlinecite{icon}.  Daniel Garber and Kurt Smith assisted the search for the source of the quotation attributed to Descartes in Ref.~\onlinecite{feet}.  Rob Iliffe kindly provided advance copy of the relevant chapters in Ref.~\onlinecite{Iliffe}.  Andrew Gray and Bernard Juby gave expert heraldic advice (Ref.~\onlinecite{Noel}).  I also thank Jos\'e Cayasso for producing Figs.~\ref{fig:Cambridge} and \ref{fig:Barrs}, Tom Hayes for help procuring titles from the Harvard libraries, Jordy Geerlings for correspondence on Hatzfeld's anti-Newtonianism (Ref.~\onlinecite{antiNewton}), Don Campbell for discussions about heraldic marks of bastardy (Ref.~\onlinecite{bastard}), as well as Ari Belenkiy and Vanessa L\'opez for general feedback.  Finally, thanks are due to the National Portrait Gallery's Rights and Images Department for permission to use \Fig{fig:Seeman}, and to Ken Redmore of the Society for Lincolnshire History and Archaeology for permission to use \Fig{fig:Roubillac}.

\end{acknowledgements}


\bibliographystyle{aipprocl}   

\begin{thebibliography}{99}
\bibitem{Euler}
Classical mechanics as it is taught today derives largely from the efforts of Euler, Lagrange, and other Continental scientists to deploy Leibniz's analytical techniques in order to facilitate and generalize the application of Newton's physical principles.  This development ---following as it did the bitter and ideologically charged disputes among Cartesians, Leibnizians, and Newtonians--- seems to me inexplicable except by its success in bringing a wide range of natural phenomena into a coherent and predictive mathematical account: i.e, by its ability to fulfill its explicit objective.
\bibitem{Bacon}
Newton is frequently identified as the first scientist knighted by the British monarch (see, e.g., Ref.~\onlinecite{Hawking}).  Yet Francis Bacon, an early advocate of experimental science, had been knighted in 1603.  Christopher Wren, sometime Savilian Professor of Astronomy at Oxford, was knighted in 1673.  Bacon held high political offices and Wren was an eminent architect, so that their knighthoods probably had little to do with their scientific pursuits.  But Newton's knighthood was probably not intended as recognition of his science either.
\bibitem{Bart}
``Bart.'' is the traditional abbreviation used after the name of a baronet.  Baronetcies are the only hereditary honors granted by the British Crown that are not titles of nobility.  Both knights and baronets are styled ``Sir,'' but knighthood is not inheritable.
\bibitem{yeoman}
A yeoman owned the land that he farmed.  According to the prevailing conventions, he ranked above laborers, artisans, and husbandmen, who owned little or no land of their own, and below gentlemen, who owned enough land to live on the rent thereof, without having to work at all.
\bibitem{notablet}
Curiously, Ref.~\onlinecite{Turnor} mentions the 1798 restoration of Woolsthorpe Manor and the plaque installed at that time in the room where Newton was born (p.\ 156), and is otherwise scrupulous in documenting local inscriptions and monuments, yet fails to mention the tablet of \Fig{fig:Woolsthorpe}.  Did the younger Mr.\ Turnor have qualms about the chirality of the coat of arms depicted there?
\end{thebibliography}

\end{document}